\documentclass[12pt,preprint]{aastex}

\newcommand{\ms}{\mbox{m s$^{-1}$}}
\newcommand{\degrees}{\mbox{$^\mathrm{o}$}}

\begin{document}

\title{On Signatures of Atmospheric Features in Thermal Phase Curves
  of Hot Jupiters}

\author{Emily Rauscher, Kristen Menou,}

\affil{Department of Astronomy, Columbia University,\\ 550 W. 120th
  Street, New York, NY 10027, USA}

\author{James Y-K.\ Cho,} 

\affil{Astronomy Unit, School of
Mathematical Sciences, Queen Mary, University of London, \\ Mile End
Road, London E1 4NS, UK}

\author{Sara Seager}
 
\affil{Dept. of Earth, Atmospheric, and Planetary sciences, and
Dept. of Physics, Massachusetts Institute of Technology, 54-1626, 77
Massachusetts Ave., Cambridge, MA, 02139, USA}

\and 

\author{Bradley M.\ S.\ Hansen,}

\affil{Department of Physics and Astronomy and Institute for
  Geophysics and Planetary Physics, University of California,
  475 Portola Plaza, Box 951547, Los Angeles, CA 90095, USA}

\begin{abstract}
  Turbulence is ubiquitous in Solar System planetary atmospheres. In
  hot Jupiter atmospheres, the combination of moderately slow rotation
  and thick pressure scale height may result in dynamical weather
  structures with {unusually} large, planetary-size scales.  Using
  equivalent-barotropic, turbulent circulation models, we illustrate
  how such structures can generate a variety of features in the
  thermal phase curves of hot Jupiters, including phase shifts and
  deviations from periodicity. {Such features may have been spotted in
  the recent infrared phase curve of HD~189733b.}  Despite inherent
  difficulties with the interpretation of disk-integrated quantities,
  phase curves promise to offer unique constraints on the nature of
  the circulation regime present on hot Jupiters.
\end{abstract}


\section{Introduction}

Just over two years ago the \emph{Spitzer Space Telescope} was used
for the first direct detections of light from extrasolar planets
\citep{Charbonneau2005,Deming2005}, heralding the birth of a new field
of observational astronomy: the characterization of exoplanet
atmospheres.  So far these direct measurements are only possible for
the shortest period ``hot Jupiter'' exoplanets, as these planets are
heated enough to radiate $\sim 10^{-3}$ of their parent stars'
infrared emission, making {the planet's component of the} signal just
separable from the stellar component.  In the last couple {of} years,
many new results have emerged in this field, including several more
infrared direct detections of hot Jupiters (and a Neptune!)  through
secondary eclipse measurements
\citep{Deming2006,Deming2007,Harrington2007}, measurements of infrared
emission spectra \citep{Grillmair2007,Richardson2007}, and orbital
phase variation from the day-night temperature contrasts of eclipsing
and non-eclipsing planets
\citep{Harrington2006,Cowan2007,Knutson2007}.  Further exciting
results are expected from the last round of cryogenic \emph{Spitzer}
observations, the ``warm \emph{Spitzer}'' phase \citep{Deming2007b},
and continuing on with the vastly increased photometric sensitivity of
the \emph{James Webb Space Telescope} ({\it JWST}).  The launch of
\emph{JWST} will enable high precision measurements and the
acquisition of detailed information about the atmospheres of these
planets, including the ability to use the known geometry of eclipsing
systems to make resolved maps of their day sides
\citep{Williams2006,Rauscher2007b}.

As the number of directly observed systems continues to grow, it {is
becoming} increasingly apparent that not all hot Jupiters are alike.
Differences in planetary radii, brightness temperatures or inferred
efficiencies of heat redistribution from permanent day sides to cold
night sides have led to theoretical speculations on possible ways to
subclassify hot Jupiters \citep{Burrows2007,Fortney2007a,Hansen2007b}.
Perhaps we should not be surprised to discover diversity in these
systems \citep[see, e.g.,][]{Redfield2007} since even our own Solar
System planets exhibit a wide range of atmospheric conditions.  For
example, it is currently not well understood what sets the wind speeds
at cloud-deck level on the Solar System gas giants.  Since atmospheric
circulation can be largely driven by insolation, one might expect those
planets closer to the Sun to experience stronger winds, but the fact
that Neptune receives much less solar illumination than Jupiter and
yet has faster winds illustrates how misleading simple arguments may
be when applied to these complex, nonlinear flows (\citealt{Cho2006};
see review by \citealt{Showman2007}).

Even a thorough understanding of the atmospheric dynamics of the
planets in the Solar System would not necessarily allow us to
extrapolate to hot Jupiters, {which are characterized by a
very} different atmospheric regime.  These presumably tidally locked,
short-period planets experience intense stellar irradiation on their
constant day sides, {much in excess of the internally generated
heat flux}, while their night sides remain in perpetual shadow.  In
addition, based on their relatively slow rotation periods, dynamical
arguments suggest that large scale atmospheric structures can form on
these planets \citep{Cho2003,Menou2003,Showman2002}, in contrast to
the many thin jets and smaller structures observed on Jupiter, for
example.  Given this unusual combination of asymmetric heating and
slow rotation, there have been several recent attempts to model the
atmospheric flow on hot Jupiters \citep[see review by Showman et
al. 2007]{Showman2002,Cho2003,Cho2006,C&S05,Cooper2006,Fortney2006,Langton2007,Dobbs-Dixon2007}.

The combination of strong external irradiation (which promotes
vertical static stability), moderately slow rotation and large
pressure scale heights in hot Jupiter atmospheres results in Rossby
deformation radii comparable to or larger than the planetary radii
themselves \citep{Showman2002,Cho2003,Menou2003,Showman2007}.
{The Rossby deformation radius acts as a limiting scale for
dynamics in a rotating stratified flow. It sets the typical horizontal
size of large vortices in a turbulent atmosphere and determines the
scale of the most unstable baroclinic modes. The large deformation
radii in hot Jupiter atmospheres may thus limit the importance of
baroclinic instabilities and favor a largely barotropic\footnote{In
this context, barotropic refers to a flow in which baroclinic
instabilities, which result in a form of horizontal convection, play a
minimal role (see \citet{Showman2007} for details).} atmospheric
flow.}

These considerations led \citet{Cho2003,Cho2006} to apply the
two-dimensional, equivalent-barotropic formulation of the primitive
equations of meteorology \citep{Salby1989} to hot Jupiter atmospheres.
{This vertically integrated formulation emphasizes the
quasi-two dimensional, barotropic nature of the flow.  While only
adiabatic simulations have been reported so far, a great advantage of
the equivalent-barotropic model is that small horizontal scales are
resolved in a fully turbulent flow \citep[see, e.g.,][for turbulent
energy spectra]{Cho1996}.  An additional advantage of this approach is
that the set of physical and numerical model parameters is limited and
it includes the unknown magnitude of global average wind speed which
is treated as an input parameter \citep{Cho2003,Cho2006}.}  This
allows us to {perform an analysis in which we explore a large range of
possible equivalent-barotropic flows for hot Jupiter atmospheres and
study how these diverse flows may translate into a diversity of
thermal phase curves for hot Jupiters.}

We begin by describing the models used and our method for computing
phase curves in $\S$~2, move on to discuss predicted model phase
curves in $\S$~3, consider implications for current and future
observations in $\S$~4, and finally conclude in $\S$~5.

\section{Method}

We use the adiabatic, equivalent-barotropic model of
\citet{Cho2003,Cho2006} for the planet HD~209458b throughout our
analysis. Our main results for thermal phase curves do not strongly
depend on the specific planetary parameters adopted---as long
as one is concerned with hot Jupiters of the same, broad dynamical
class as HD~209458b \citep[see, e.g.,][]{Menou2003}.  The reader is
referred to \citet{Cho2006} for a detailed description of the
circulation model.  Briefly, the stellar radiative forcing imposes a
steady day-night temperature difference on the flow.  In the adiabatic
equivalent-barotropic model, this forcing is represented by a forced
bending of the model's bottom boundary surface, with a specified bend
amplitude $\eta$. {This corresponds to a maximum day-night
temperature variation of amplitude $\Delta T=\pm \eta \bar{T}$, where
$\bar{T}$ is the average equivalent temperature of the modeled layer.
The pre-existing background flow, with a prescribed average global
wind speed, $\bar{U}$, produces temperature variations, which are
superimposed on those due to the steady day-night difference.  The
combined temperature range covered by each model, as well as the range
due to radiative forcing only (case $\bar{U}=0$), are given in
Table~1.}  The model runs have been performed at T63 spectral
resolution (equivalent to a fully-resolved 192 $\times$ 96
longitude-latitude grid), which has been found to be sufficient to
capture the formation of dominant atmospheric structures
\citep{Rauscher2007}.  Following the initialization procedure
described in \citet{Cho2003,Cho2006}, each run {has been} allowed to
relax for 40 {planetary} orbits, after which we output a snapshot of
the temperature structure on the planet, 100 times per orbit, for the
next 60 orbits.

We create the phase curves presented here by rotating each temperature
map to the appropriate viewing orientation for its orbital phase angle
and assumed inclination, and orthographically projecting it onto a two
dimensional disk.  Unless otherwise specified, a $90$\degrees
~inclination (edge-on orientation) is assumed by default.  We sample
the disk with a grid of $[r=20,\theta=40]$ resolution.  We solve for
three values of emergent flux from each disk element area: a
bolometric value $\propto T^4$, and \emph{Spitzer} band fluxes
calculated either from a blackbody spectrum based on the local
temperature $T$ or a spectral model interpolated from a grid of
cloudless 1-D radiative transfer calculations \citep{Seager2005}. As
we shall see, our main conclusions depend only weakly on the specific
emission model adopted.  {After obtaining the flux}, we integrate over
the disk, weighting the emergent flux from each grid point by its
relative area on the disk.  Our procedure is similar to that adopted
in \citet{Rauscher2007}.

We performed resolution tests of our phase curve analysis, both in the
number of temperature maps sampled per orbit and the number of grid
elements used to sample each map.  For computational speed, we use
$[r=20,\theta=40]$ as the lowest possible resolution that is immune to
noise from under-sampling.  The phase curves produced using this
resolution are accurate to within a few percent of those produced
at higher resolutions.  Similarly, we find that by sampling the
temperature map of the planet 50 times per orbit, we are able to
adequately track the orbital flux variation without losing details in
the structure of the phase curve.  We have also found that 10 orbits
are sufficient to sample the variable patterns seen in the phase
curves, making it unnecessary to analyze the full 60 orbits of our
model outputs.

\section{Phase Curves from Equivalent-Barotropic Models}

The temperature structures predicted by {the equivalent-barotropic
model} consist of two superimposed components: a global day-night
``fixed'' temperature gradient {(the forced bending---see \S~2)} whose
strength depends on the imposed radiative forcing (via the parameter
$\eta$), and a variable pattern of weather features created by
atmospheric winds.  The most prominent weather features are cold
circumpolar vortices whose strength {(vorticity)} increases with
higher values of $\bar{U}$.  The vortices rotate around the poles with
a period comparable, but not equal, to the spin {(=orbital)} period of
the planet.  While these vortices are the weather features that most
strongly affect the emitted thermal flux in our models, other large
scale, possibly transient, features generally exist in {an}
atmosphere.  For example, in the models presented here we observe a
broader warm area that develops on the side of the planet opposite
from the circumpolar vortices, and on occasion cold transient features
that develop on the night side.  {These features constitute the
weather on these model planets.}  Figure~\ref{fig:images} shows
temperature snapshots for the $\eta=0.10$, $\bar{U}$ = 800 \ms~model,
exemplifying the type of weather features that develop.  Our
discussion will primarily focus on the circumpolar vortices, as they
are the {dynamical} features that most strongly affect the orbital
phase curves, but it is worth noting that they do not completely
determine the observable signatures in the model phase curves.

The nature of the global temperature structure is determined by the
relative strengths of the two superimposed components mentioned above.
The models with high $\eta$ and low $\bar{U}$ values have strong
day-night temperature contrasts with weak variable weather patterns.
Their thermal emission phase curves are thus primarily a simple
day-night (peak-trough) pattern with only minor deviations from weak
weather features.  The low $\eta$, high $\bar{U}$ models, on the other
hand, retain some day-night temperature difference, but they can have
temperature structures strongly influenced by the variable weather
features.  The corresponding phase curves are highly variable from one
orbit to the next, {since they are strongly dependent on the
location of cold polar vortices relative to the day-night gradient.}
Here we survey the $(\eta, \bar{U})$ parameter space to determine the
range of possible characteristics of model phase curves.

\subsection{The Influence of Wind Strength} \label{sec:ubar}

\begin{figure}[ht]
\begin{center}
\includegraphics[width=\textwidth]{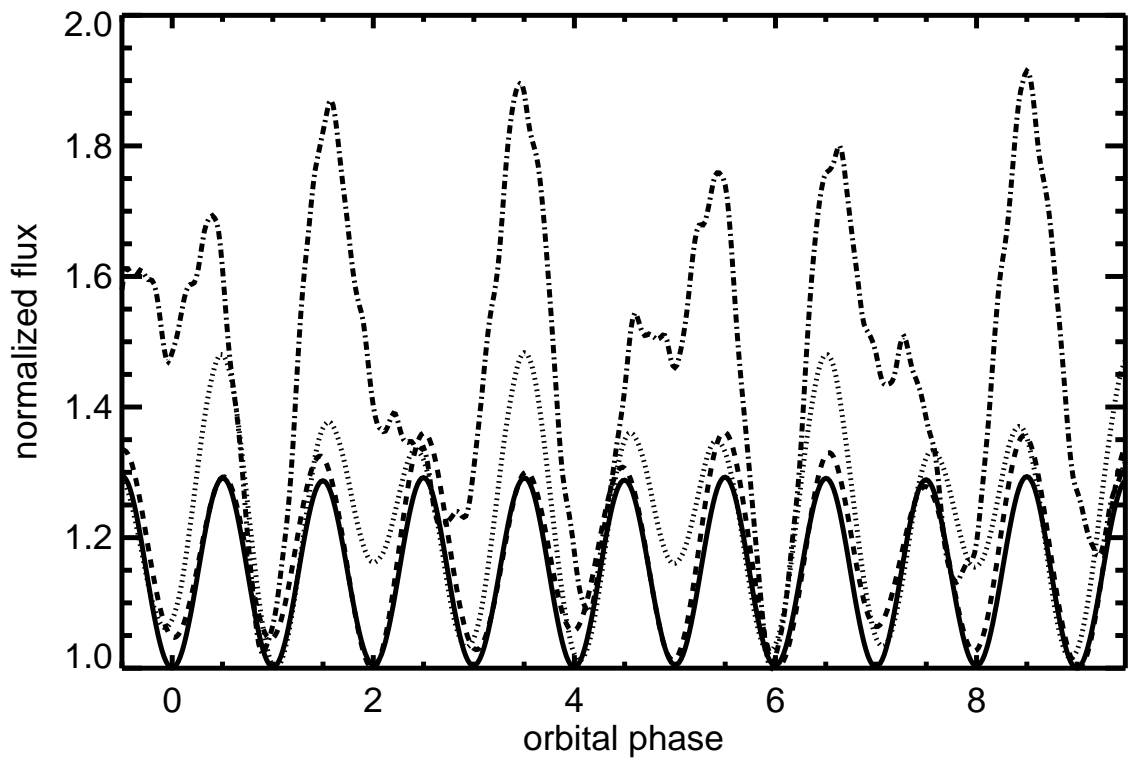}
  \caption{Bolometric orbital phase curves for equivalent-barotropic
    model {simulations} with $\eta=0.05$ and $\bar{U} =$ 800
    (\emph{dash-dotted}), 400 (\emph{dotted}), 200 (\emph{dashed}),
    and 100 (\emph{solid}) \ms.  Each curve is normalized to its
    absolute minimum, emphasizing the relative variation for each
    model.  The phase curves become increasingly complex in models
    with stronger winds because the temperature field is increasingly
    dominated by variable weather features.  Observable signatures of
    atmospheric features include small {perturbations} in the phase curve,
    offsets in the phase of min./max. flux, and variations in the peak
    to trough amplitude, from one orbit to the
    next.}\label{fig:u_values}
\end{center}
\end{figure}

First we consider models with the radiative forcing parameter $\eta$
fixed ($\eta=0.05$).  Figure~\ref{fig:u_values} shows the bolometric
phase curves resulting from runs with average global wind strength
$\bar{U}$ ranging from 100 to 800 \ms.  One immediately apparent
result is that the amount of predicted flux variation increases with
$\bar{U}$.  This is to be expected, since rotating cold polar vortices
and other weather features have an increasingly dominant role in the
temperature structures of models with stronger wind strengths.  For
example, when the vortices are on the night side, there is a stronger
day-night temperature difference than would result from radiative
forcing alone, and this allows for more orbital flux variation than in
models where the vortex-induced temperatures are weak compared to
the imposed day-night temperature difference.

We do not consider global wind strengths in excess of 800 \ms \,in
this study. It is unclear whether much larger wind speeds
\citep[$\sim$ several km/s; e.g.,][]{C&S05,Dobbs-Dixon2007} can be
realized in a turbulent, rotationally-balanced flow on hot
Jupiters. We have found that large wind speeds of a few km/s are not
possible in our specific setup since they either lead to atmospheric
holes in the initially balanced state or to a rapid blow up of the
flow field \citep{Cho2006}.

It is worth noting that in these adiabatic models the strength of the
imposed day-night temperature is decoupled from the assumed global
mean wind strength: it depends solely on $\eta$. {Winds of
greater strength} may cause a decreased day-night temperature
difference by, for example, more efficiently redistributing the heat
between the two sides.  This means that the trend of increasing phase
curve amplitude with $\bar{U}$ shown here may not be self-consistent.
Coupled, diabatic circulation models would address this issue more
satisfactorily.

The next feature that emerges in {cases} with higher $\bar{U}$ is the
presence of {oscillations} in the phase curves.  Whenever the cold
vortices are visible on the planetary disk, they create a dip in the
disk-integrated thermal flux.  These dips can weaken the day side flux
peak or strengthen the night side flux trough, when the vortices are
predominantly at the sub- or anti-stellar point, respectively.  The
vortices and other weather features also produce small {perturbations} seen in
the phase curves (most apparent for the $\bar{U}=800$ \ms~model), when
apparent at some intermediate phase between peak and trough.

The smallest of the identifiable {perturbations} is about 30\degrees~wide in
orbital phase.  It is not trivial to relate the phase size of this
{perturbation} back to a physical scale in the planet's {atmosphere}, since
there exists some degeneracy in interpretation between the size and
brightness of any atmospheric feature.  {F}or instance, a small and
extremely bright feature that {is present during the entire passage
across the visible disk could result} in a phase curve {perturbation} of
almost 180\degrees.  If the feature were dimmer, perhaps it would only
significantly contribute to the observed flux when it was near the
``sub-observer'' point, and the size of the resulting {perturbation} would
therefore be limited in phase.  However, a diffuse feature of only
slightly elevated brightness could produce a similar {perturbation}.  Thus the
size and brightness of features can counterbalance each other, making
it difficult to place constraints on either property.  This
exemplifies the ambiguity present in any interpretation of orbital
phase curve data---each point in the orbit is a measure of the
disk-integrated emission so that, in principle, it is possible to
construct a variety of temperature structures that reproduce a given
orbital variation.

In our model {runs}, when the circumpolar vortices are close to being
aligned with the substellar point, but are some degrees of longitude
displaced, the peak of the emitted phase curve is somewhat offset from
the phase at which the substellar point faces the observer.  The
resulting apparent phase shift could be interpreted as a hot spot
advected away from the substellar point, while in this example it is
actually the result of the disk-integrated combination of emission
from the hot substellar region and the asymmetry from the cold
pattern.  This point is worth emphasizing, as it shows that there can
be degeneracy between models in the interpretation of even a single
orbit's phase curve.  Examining Figure~\ref{fig:u_values} we find
phase offsets of up to $\pm$40\degrees~(with extreme offsets of orbit
8 at 85\degrees~and orbit 3 at 100\degrees) for the $\bar{U}=800$
\ms~model {run}.  The $\bar{U}=400$ and 200 \ms~model {runs} typically
produce phase shifts of $\pm$30\degrees~and $\pm$10\degrees,
respectively.

\subsection{The Influence of Imposed Radiative Forcing}

\begin{figure}[ht]
\begin{center}
\includegraphics[width=\textwidth]{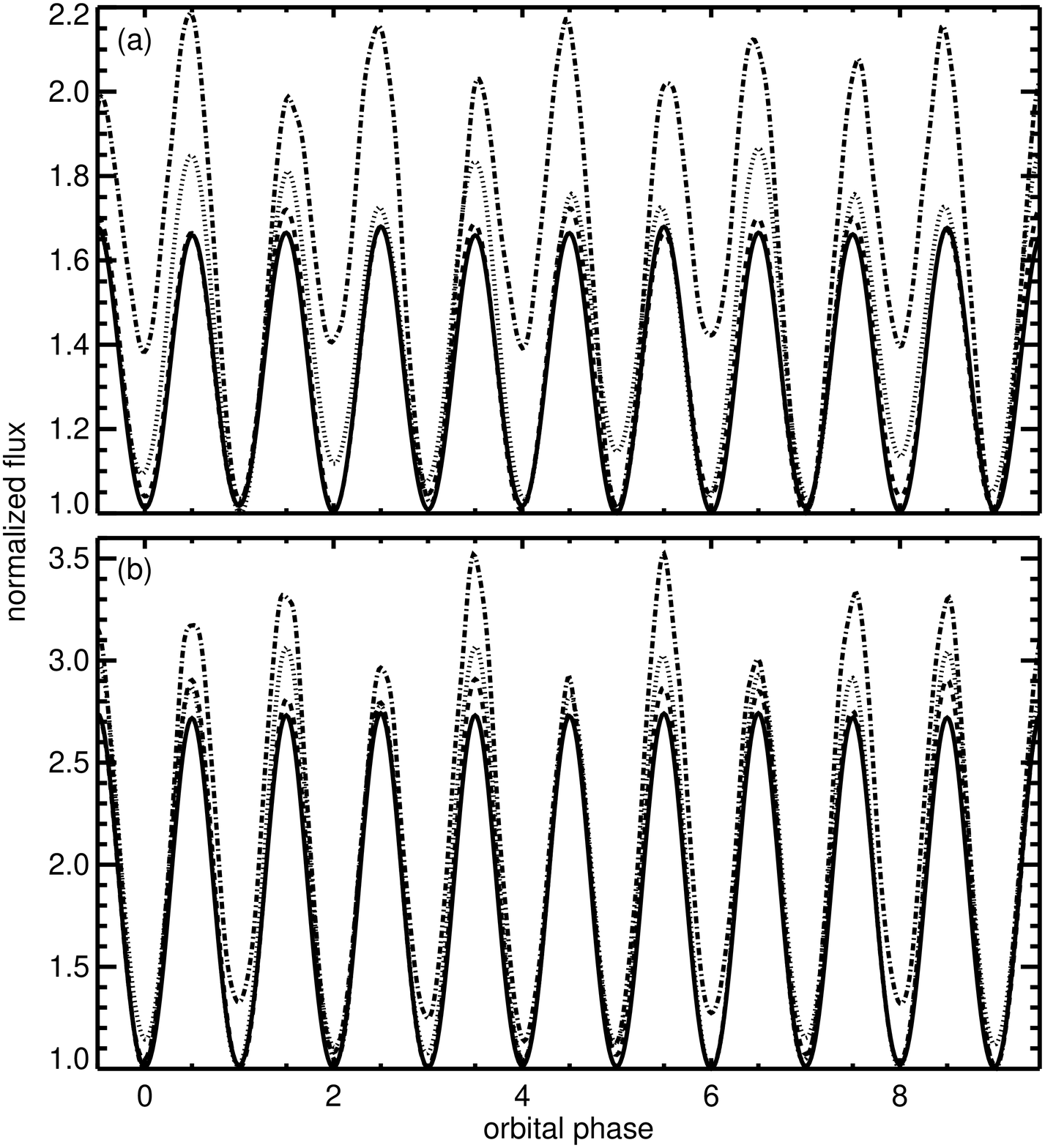}
  \caption{Same as Figure~\ref{fig:u_values} (from top to bottom:
    $\bar{U}=800, 400, 200, 100$ \ms), but for {simulations} with
    $\eta=0.10$ (\emph{a}) and $\eta=0.20$ (\emph{b}).  As the amount
    of radiative forcing is increased, the imposed day-night
    temperature contrast becomes the dominant component of the
    planet's temperature field, rendering the phase curves more
    regular in their variation.  Nevertheless, large-scale weather
    features continue to produce deviations away from a strictly
    regular peak-trough pattern, even for the strong radiative forcing
    case.} \label{fig:eta_values}
\end{center}
\end{figure}

Next we consider the effect of increasing the amount of radiative
forcing imposed on the model atmospheres, controlled by the parameter
$\eta$.  Figure~\ref{fig:eta_values} shows phase curves for the full
range of $\bar{U}$ values for {simulations} with $\eta = 0.10$ and
0.20.  As expected, the increased radiative forcing causes the regular
day-night variation to dominate over the features produced by the
variable weather structure.  The day-night variation is completely
dominant for the weakest wind models; for $\eta=0.20$, a wind speed of
800 \ms~is required to produce deviations of more than a few percent
away from the peak-trough pattern.  Therefore, increasing the value of
$\eta$ results in phase curves with larger amplitudes and a more
regular peak-trough pattern.  The cold vortices are too weak, compared
to the day-night gradient, to produce the same small features seen in
the phase curves of the $\eta=0.05$ models, and they now only work to
change the amplitude of the variation from one orbit to the next, or
to produce small shifts in the phase of the peak flux.  The amount of
phase offset is greatly reduced from the $\eta=0.05$ model. For
$\eta=0.10$, $\bar{U}=800$ \ms~leads to $\pm$20\degrees ~offsets,
$\bar{U}=400$ \ms~leads to $\pm$10\degrees ~offsets and there are no
shifts for $\bar{U}=200$ \ms~or less. For $\eta=0.20$, the only model
whose curve shows a shift is $\bar{U}=800$ \ms, with an offset of
$\pm$10\degrees.

\subsection{The Role of Inclination}

\begin{figure}[ht]
\begin{center}
\includegraphics[width=\textwidth]{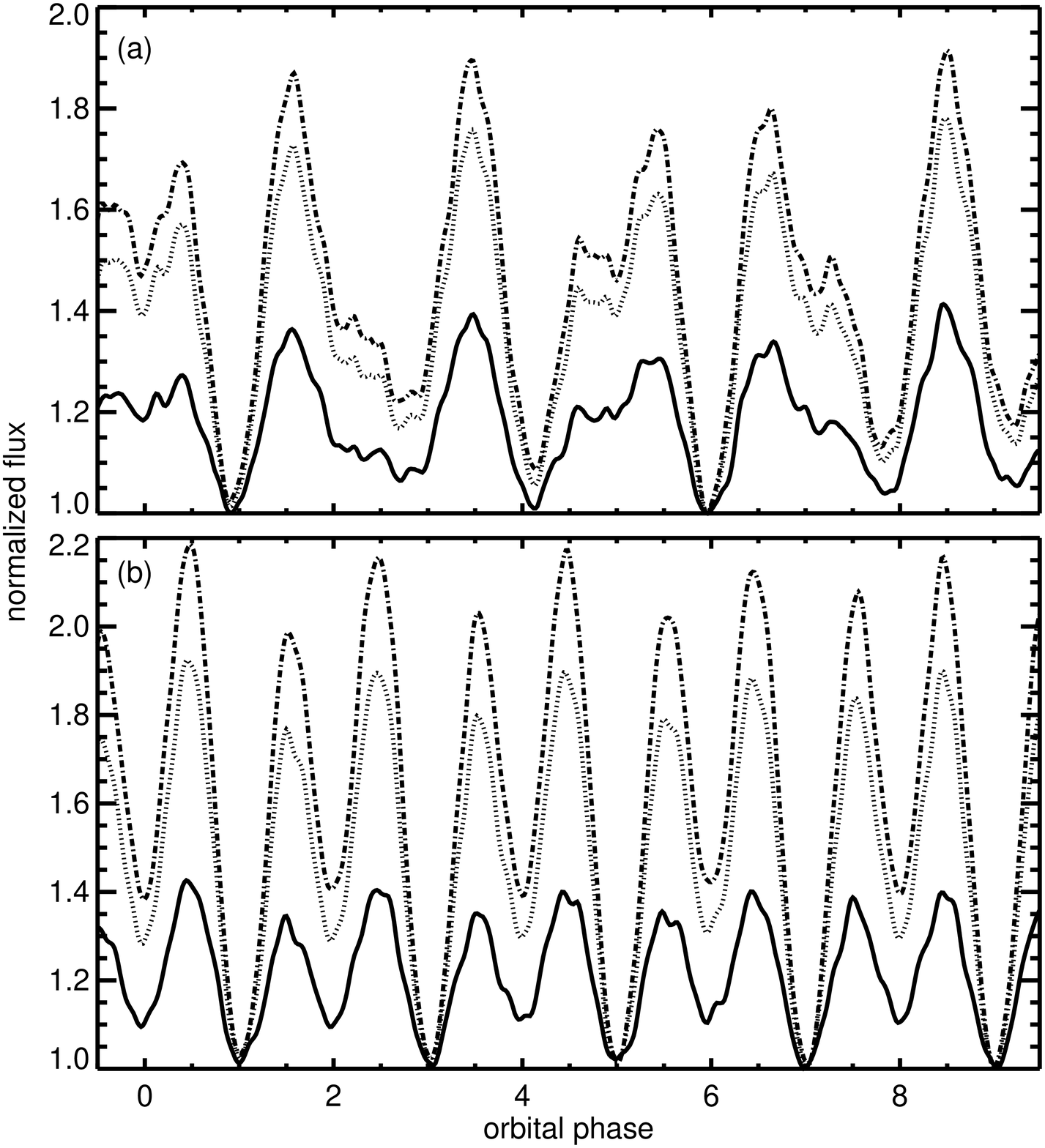}
  \caption{\small Orbital phase curves for $\bar{U}=800$
    \ms~equivalent-barotropic models with $\eta=0.05$ (\emph{a}) and
    $\eta=0.10$ (\emph{b}). Three cases, with orbital inclinations $i
    =$ 90\degrees~(edge-on; \emph{dash-dot}),
    60\degrees~(\emph{dotted}), and 30\degrees~(\emph{solid}), are
    shown in each panel.  As the system's inclination is tilted away
    from edge-on {orientation}, the amount of phase variation due to
    the imposed day-night temperature contrast is decreased, while the
    signatures of variable weather features become increasingly more
    apparent in the phase curves.  The features that contribute most
    strongly to the wiggly shapes of these model phase curves are the
    cold circumpolar vortices, since the change in inclination brings
    them more into view.} \label{fig:incl_values}
\end{center}
\end{figure}

In addition to sampling the $(\eta,\bar{U})$ parameter space itself,
in this work we also consider the role of orbital inclination on model
phase curves.  A challenge in interpreting data from non-eclipsing
systems is that the unknown inclination is partially degenerate with
the amount of heat redistributed to the night side
\citep[e.g.,][]{Harrington2006}.  In Figure~\ref{fig:incl_values} we
show the effect of varying the assumed inclination in our calculations
of the phase curves for model {runs} with $\eta=0.05$ and 0.10, and
$\bar{U}=800$ \ms.  We choose to plot the highest wind {runs} (and
likewise omit the $\eta=0.20$ {run}) {since} the role of inclination
on phase curves from atmospheres dominated by the day-night
temperature contrast is simple: as the inclination is reduced (i.e.,
as the orbit is viewed from an increasingly polar orientation), there
is a trivial decrease in the amplitude of orbital phase variation.
(At the $i=0$\degrees~extreme there is no variation since equal parts
of the day and night sides are constantly in view.)  We see this
reduction in amplitude in our model curves as well, especially for the
$\eta=0.10$ model, where the thermal emission is dominated by the
day-night temperature contrast.

\begin{figure}[ht]
\begin{center}
\includegraphics[width=\textwidth]{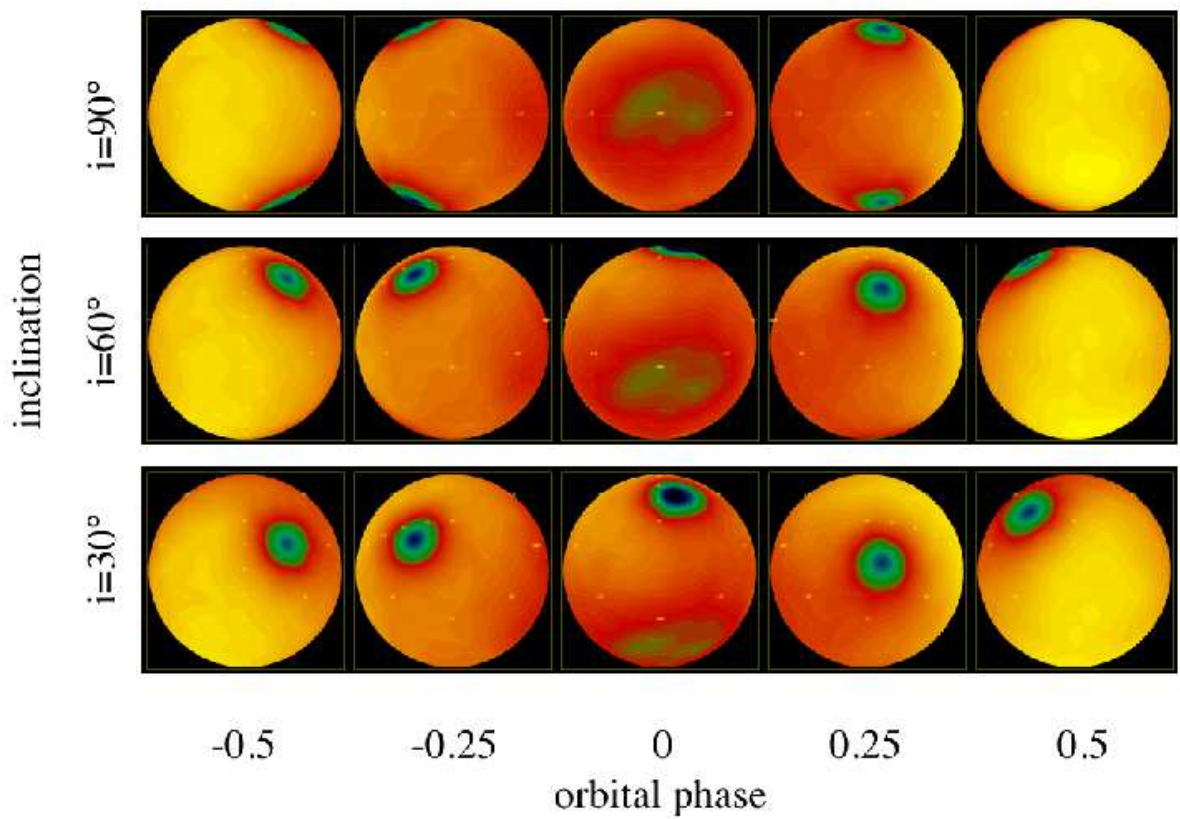}
  \caption{Snapshots of the apparent temperature field in the model
    with $\eta=0.10$, $\bar{U}=800$ \ms,~as it rotates through one
    orbit.  Temperatures range from 700 to 1400K.  From left to right,
    the central longitudes are: 0\degrees~(substellar; phase $-0.5$),
    90\degrees, 180\degrees~(antistellar; phase $0$), 270\degrees, and
    360\degrees.  The maps in the top row are shown at an inclination
    of 90\degrees~(edge-on), the middle ones at 60\degrees~and the
    bottom ones at 30\degrees.  As the inclination of the system is
    titled away from edge-on {orientation}, the cold
    circumpolar vortices assume a more central location on the
    apparent planetary disk, so that they affect more strongly the
    disk-integrated thermal emission.  At the same time, more equal
    areas of the hot day side and cool night side are visible, which
    decreases the amplitude of the regular peak-trough pattern in the
    phase curves. \emph{[See the electronic edition of the Journal for
      a color version of this figure.]}}\label{fig:images}
\end{center}
\end{figure}

Another effect of decreasing the inclination is the emergence of more
small features in the model phase curves.  Since the cold vortices
revolve around the poles of the planet, as the orbital inclination is
decreased, they are located closer to the center of the observed
planetary disk, and therefore are given more weighting in the
disk-integrated emitted flux.  For a visual explanation of this
effect, see Figure~\ref{fig:images}, where we have plotted temperature
snapshots from one orbital rotation of the $\eta=0.10$, $\bar{U}=800$
\ms~model {run}, shown at $i =$ 90, 60, and 30\degrees.  Here we show
the $\eta=0.10$ model {run} because it has strong enough radiative
forcing that the day-night gradient is clearly seen, while
$\bar{U}=800$ \ms~creates strong enough vortices that they are
likewise easily visible in the global temperature maps.  The top,
middle, and bottom rows of Figure~\ref{fig:images} correspond to the
top, middle, and bottom curves of Figure~\ref{fig:incl_values}b,
respectively.  We see that at $i=30$\degrees, the cold vortices are
more centrally located on the apparent disk and the day-night
temperature difference plays a much less significant role in producing
variations of the disk-integrated flux.  The decrease in day-night
variation, combined with the more prominent role of cold vortices,
results in the increased presence of small features in the model phase
curve.

\subsection{Wavelength-Dependent Phase Curves}

\begin{figure}[ht]
\begin{center}
\includegraphics[width=\textwidth]{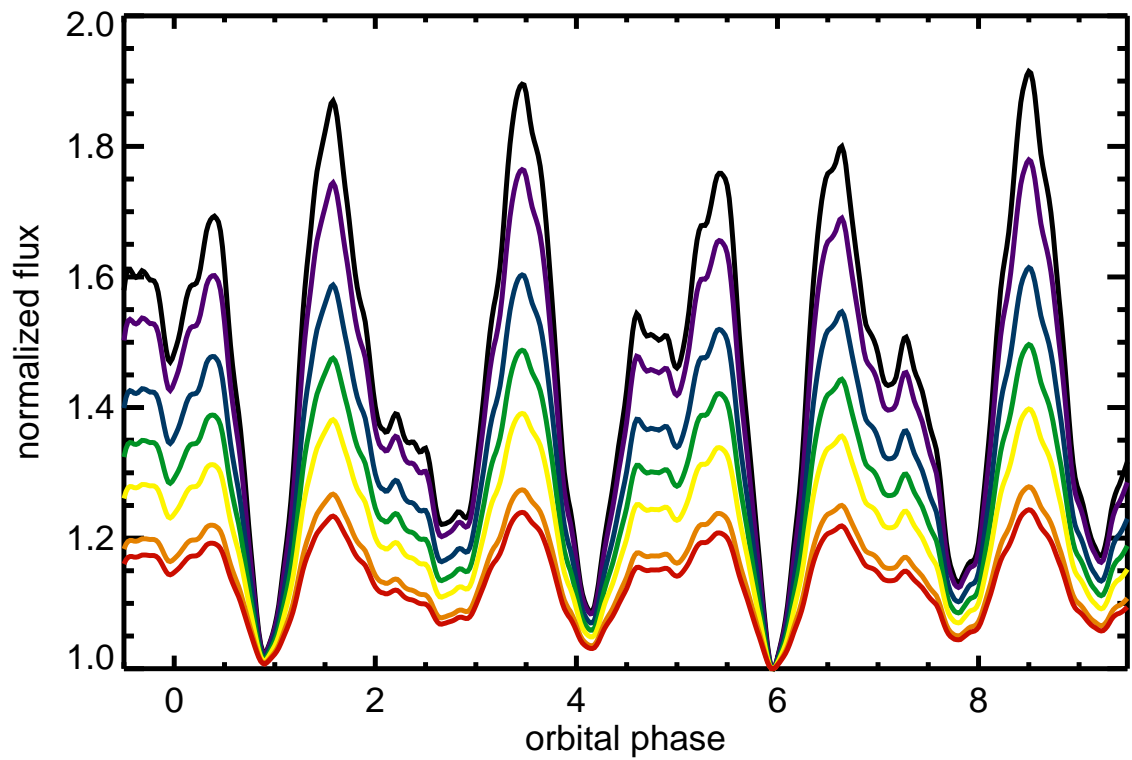}
  \caption{Orbital phase curves as they would be observed in various
    \emph{Spitzer} instrumental bands, for the $\eta=0.05$,
    $\bar{U}=800$ \ms~equivalent-barotropic model, assuming simple
    blackbody emission.  From top to bottom the curves are: bolometric
    (\emph{black}), 3.6\micron~(\emph{purple}),
    4.5\micron~(\emph{blue}), 6\micron~(\emph{green}),
    8\micron~(\emph{yellow}), 16\micron~(\emph{orange}), and
    24\micron~(\emph{red}).  The shorter wavelengths are more variable
    for the range of temperatures covered by this model.  \emph{[See
      the electronic edition of the Journal for a color version of
      this figure.]}}
\label{fig:bbody}
\end{center}
\end{figure}

Throughout our discussion so far we have been analyzing bolometric
phase curves---\emph{i.e.}, light curves calculated under the simple
assumption that the emission from any area on the planetary disk is
$\propto T^4_{local}$.  {Here} we also compute the light curves that
would be measured in each \emph{Spitzer} instrumental band to
determine the wavelength dependence of these model curves.  We
calculate two values for the flux measured by each \emph{Spitzer}
band: one uses a blackbody for the emitted spectrum from each element
area on the planetary disk, and the other one interpolates from a grid
of model atmosphere spectra from \citet{Seager2005}. This is identical
to the procedure adopted in \citet{Rauscher2007,Rauscher2007b}.
Figure~\ref{fig:bbody} shows the wavelength-dependent phase curves for
the $\eta=0.05$, $\bar{U}=800$ \ms~{simulation}, assuming local
blackbody emission.  The amplitude of variation decreases with
increasing wavelength.  {We have verified that, in the range of
temperatures covered by our model {runs} (see Table~1), this is simply
a consequence of the Planck function varying more with temperature at
shorter wavelengths than it does at longer wavelengths.}

\begin{figure}[ht]
\begin{center}
\includegraphics[width=\textwidth]{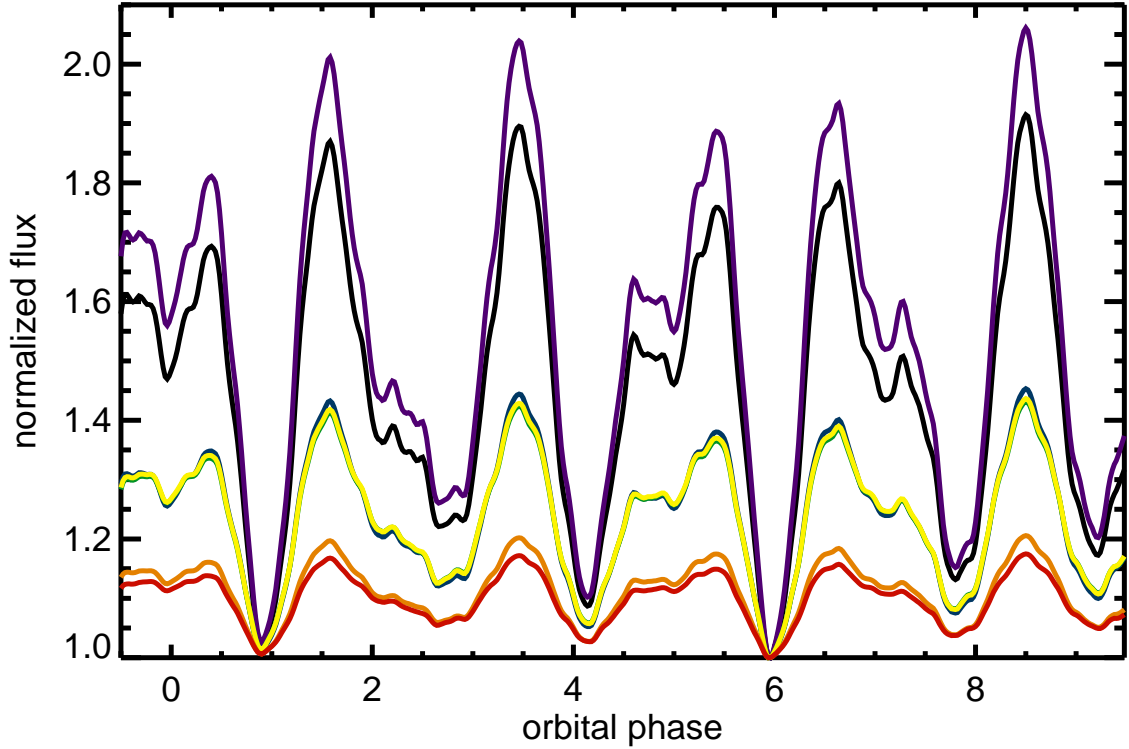}
  \caption{Orbital phase curves as they would be observed in various
    \emph{Spitzer} instrumental bands, for the $\eta=0.05$,
    $\bar{U}=800$ \ms~equivalent-barotropic model, using detailed
    atmospheric spectral models for emission.  From top to bottom the
    curves are: 3.6\micron~(\emph{purple}), bolometric (\emph{black}),
    followed by overlapping 4.5\micron~(\emph{blue}),
    6\micron~(\emph{green}), and 8\micron~(\emph{yellow}), and finally
    16\micron~(\emph{orange}) and 24\micron~(\emph{red}). The results
    for any given band are strongly affected by the absorption
    properties of constituents included in the atmospheric spectral
    model (especially water), but shorter wavelengths are generally
    more variable for the range of temperatures covered by our
    circulation model.  \emph{[See the electronic edition of the
      Journal for a color version of this figure.]}}
\label{fig:smodel}
\end{center}
\end{figure}

This same general behavior is seen in Figure~\ref{fig:smodel}, in
which we plot model phase curves for the same \emph{Spitzer} bands,
but using the detailed spectral models for local emission.  Just as
with a blackbody, the modeled spectra are more temperature-sensitive
at shorter wavelengths.  The detailed wavelength-temperature
dependence of these spectral models differs from that of a blackbody,
however, largely because of the presence of absorption bands,
principally those due to water in the atmosphere.  Although the
presence of water {vapor spectral signatures} in hot Jupiter
atmospheres is currently debated
\citep{Barman2007,Burrows2007,Fortney2007b,Grillmair2007,Richardson2007,Tinetti2007},
the main point of Figures~\ref{fig:bbody} and~\ref{fig:smodel} remains
the same: for an atmosphere that spans the range of temperatures
expected on hot Jupiters, phase curves at shorter wavelengths may be
generally more sensitive to temperature differentials.

One needs to interpret this trend cautiously, however.  It is only one
of several effects that contribute to the detailed emission properties
from such an atmosphere.  For example, our modeling strategy based on
a vertically-integrated formulation of the atmospheric flow does not
capture well the possibility that shorter wavelengths may probe deeper
into a planet's atmosphere \citep[e.g.,][]{Seager2005}, where the
circulation regime and the horizontal temperature field could in
principle differ from higher up.  {While the model phase curves shown
in Figs.~\ref{fig:bbody} and~\ref{fig:smodel} emphasize the role of
horizontal temperature variations, additional wavelength-dependent
effects are expected from variations of temperature with atmospheric
depth.  {The vertical profile therefore has an effect on the
wavelength dependence of phase curves which is unspecified in our
models and the behavior shown here can thus only be associated with
effect of horizontal temperature variations.}  Three-dimensional
radiation-hydrodynamics models will be required to address this issue
reliably.}

\section{Implications for current and future observations of hot Jupiters}

The equivalent-barotropic circulation models of \citet{Cho2003,
Cho2006} indicate the possible existence of large scale weather
structures {on hot Jupiters.  These structures could produce
observable signatures in the thermal phase curves of hot Jupiters, in
the absence of obscuring elements such as clouds or haze \citep[see,
e.g.][]{Pont07}.}  We have surveyed the parameter space of these
models and find a diverse set of resulting phase curves.  This
diversity may be relevant in the interpretation of observations from
various systems, especially as we are beginning to find that the label
``hot Jupiter'' may in fact encompass a range of planetary types
\citep[e.g.,][]{Burrows2007,Fortney2007a,Hansen2007b}.

Even within the limited set of observed infrared phase curves, there
is already evidence for significant diversity.  The first published
exoplanet phase curve was a set of five \emph{Spitzer} 24$\micron$
observations of $\upsilon$ Andromeda b during one of its 4.617-day
orbits \citep{Harrington2006}.  These authors have found strong
variation, indicative of little to no heat redistribution, and a small
phase offset consistent with zero.  \citet{Cowan2007} used the IRAC
instrument on \emph{Spitzer} to observe HD 209458, HD 179949, and 51
Peg.  Their 8$\micron$ data placed an upper limit on the efficiency of
heat redistribution on HD 179949b, while both HD 209458b and 51 Peg b
were given lower limits.  There was no obvious phase shift evident in
the data, and these authors chose to fit their data with zero
phase-offset models.  \citet{Hansen2007a} report 24$\micron$ phase
curves, with observations at five epochs for $\upsilon$
And\footnote{These are the same data as reported in
\citet{Harrington2006}.}, $\tau$ Bootes, and 51 Peg, and at three
epochs for the fainter systems HD 179949 and HD 75289.  Like for 51
Peg, and in contrast to $\upsilon$ And, there is some evidence for
heat redistribution on $\tau$ Boo.  Particularly interesting is the
detection of an approximately -80\degrees~shift in the phase curve of
$\tau$ Boo.  This is a phase lag, meaning that the minimum of the
curve occurs after transit. {Recent MOST photometry of the
system indicates that this phase variation may be due to the presence
of an active region on the star \citep{Walker2008}. In such a case, it
would be more difficult to disentangle the variation in emission due
to any longitudinal temperature differences on the planet from
additional variations due to the presence of magnetic structures.
Nevertheless, the possibility of both positive and negative phase
offsets, as found in the phase curves presented here, and as opposed
to systematically positive ones according to other dynamical models
\citep[e.g.,][]{Showman2002,C&S05,Fortney2006,Langton2007} may provide
important additional constraints on the regime of circulation present
in these atmospheres.}

The most detailed observational results published to date are those of
\citet{Knutson2007}.  While the interpretation of other phase
observations has been limited by the necessity to fit model curves to
a few noisy points spread throughout an orbital period,
\citet{Knutson2007} observed HD 189733b continuously for $\sim$33
hours, just over half an orbital period.  These authors report an
amplitude of variation indicating heat redistribution, although at an
incomplete level.  The flux maximum occurs 2.3 hours before secondary
eclipse, while the minimum is 6.7 hours after transit (=primary
eclipse), which corresponds to two separate phase shifts: a
+16\degrees~phase shift for the flux maximum, and a
$-$47\degrees~shift for the minimum.  This means that the
data cannot be explained with a simple model that consists solely of a
day-night temperature gradient, nor a model whose only atmospheric
feature is a hot spot shifted away from the substellar point.

The continuous nature of the \citet{Knutson2007} data allows for the
resolution of small features in the phase curve, which was not
possible with other published measurements.  By postulating a steady
emission pattern over the observation time (in the frame rotating with
the synchronized planet), these authors reconstruct a longitudinal map
of the brightness across the planetary disk.  The disk-integrated flux
reaches its minimum and maximum values in the same longitudinal
hemisphere and a noticeable {small scale} feature
$\sim$20\degrees~wide in phase is seen directly after transit.

Interestingly, we find that similar {small scale perturbations}
can arise in the thermal phase curves of our equivalent-barotropic
models. In particular, the $\eta=0.05$, $\bar{U}=800$ \ms~model
(Figs.~\ref{fig:u_values} and~\ref{fig:bbody}) shows such features and
produces an overall amplitude of phase curve variations that is
qualitatively consistent with the \citet{Knutson2007} data.
{The min-to-max variation in the data is $\sim$50\%, a value
consistent with the approximate 20--80\% range of single-orbit flux
variation seen in this model.}  Since {the features of this
model phase curve} rely on the presence of large-scale moving
atmospheric features, a consequence is that the shape of the phase
curve would be expected to vary from one orbit to the next according
to our models.  This is in fact one of the main qualitative
predictions of high-resolution, equivalent-barotropic models for hot
Jupiter atmospheres: with strong weather patterns, the temporal
variability of the atmospheric temperature field should result in
possibly detectable variations for repeated observations of a single
system \citep[see also][]{Menou2003,Rauscher2007}.

{This is particularly relevant in light of the recent
24$\micron$ observation of HD 189733b \citep{Knutson2008}, in which
another continuous measurement of the system over a similar half-orbit
was performed.  Although the circulation regime (e.g., global average
wind speeds) could be different in the atmospheric layers probed at 8
and 24$\micron$ \citep[e.g.,][]{Seager2005}, the phase curve in
\citet{Knutson2008} is similar in amplitude and shape to the
8$\micron$ observation.  Thus the global temperature structure seems
to have little vertical variation between the two corresponding
``photospheres.''  Perhaps more importantly, as it pertains to our
models, there is little difference in the observed emission pattern
between these two orbits.  This disfavors any atmospheric model for
this planet that predicts significant orbit-to-orbit variations in
temperature structure, and specifically the $\eta=0.05$, $\bar{U}=800$
\ms~model previously deemed qualitatively consistent with the
8$\micron$ observations.  While the small feature seen after transit
in the 8$\micron$ data is still indicative of some complexity in the
global temperature structure, and the 24$\micron$ data are not of
sufficient precision to observe a comparable feature
\citep{Knutson2008}, a comparison between these two phase curves
points toward a mostly steady global circulation pattern on HD
189733b.}

The interpretation of any orbital phase curve data is subtle.
Measurements that consist of several discrete observations throughout
one orbit may constrain the level of heat redistribution around the
planet, but there are degeneracies even with the simplest models when
trying to fit the data \citep{Hansen2007a}.  This type of observation
may not easily constrain the level of atmospheric complexity inherent
in nonlinear circulation models---even for models as simple as the
ones used here.  The interpretation of observational data
{sets} like {those} of \citet{Knutson2007,Knutson2008}
becomes ambiguous if variable weather systems are present in hot
Jupiter atmospheres, since they can typically vary on orbital
timescales.  The assumption of a steady atmospheric field (in the
planet's rotating frame) is necessary for the construction of a
longitudinal temperature (or brightness) map, so that the resulting
inferred properties could become inaccurate if moving features exist.

Repeated observations over multiple orbits help to constrain the
steadiness of the global temperature field, as opposed to one which
changes from orbit to orbit, like the high $\bar{U}$ models presented
here.  The permanence of the day side field could also be constrained
with repeated secondary eclipse measurements \citep{Rauscher2007},
which are less observationally expensive than phase curve measurements
but lack information on the global day-night pattern.  In the future,
these two observational techniques should usefully complement each
other when applied to eclipsing systems.

Finally, while we have focused our study on thermal phase curves, it
is worth noting that the upcoming \emph{Kepler} mission could
potentially also constrain the presence of large-scale weather
features in (very) hot Jupiter atmospheres.  Indeed, signatures
similar to those in thermal phase curves may emerge in reflected light
phase curves, if dynamical weather structures such as circumpolar
vortices act as distinct thermodynamic regions of the atmospheric
flow, with specific cloud and albedo properties.  These issues would
be best addressed in the future with improved atmospheric circulation
models, coupled with radiation transfer calculations.

\section{Conclusion}

The interpretation of hot Jupiter phase curve data is subtle,
especially if variable weather features are present in their
atmospheres.  We have explored the parameter space of the adiabatic,
equivalent-barotropic models of \citet{Cho2003,Cho2006} to investigate
the possible effects that large-scale features would have on hot
Jupiter thermal phase curves.  Despite their relative simplicity, we
find that these models can produce phase curves with shapes that
change from one orbit to the next, have apparent phase shifts, and
contain {small transient perturbations}.  The detection of such
observational signatures would probe the regime of circulation present
in these atmospheres.  Even the absence of such features would
constrain the atmospheric dynamics regime.

{We thank the anonymous referee for comments that helped us
clarify the manuscript.} This work was supported by NASA contract
NNG06GF55G.

\clearpage

\begin{deluxetable}{cccc}
\tablewidth{0pt}
\tablecaption{Equivalent-barotropic models under consideration}
\tablehead{
\colhead{$\bar{U}$}  & \multicolumn{3}{c}{Min-Max Layer Temperature (K)} \\
\colhead{(\ms)}  & \colhead{$\eta=0.05$}  & \colhead{$\eta=0.10$}  & \colhead{$\eta=0.20$}
}
\startdata
0\tablenotemark{a}  &  1140-1260   &  1080-1320   &  960-1440   \\
100  &  1148-1274   &  1085-1335   &  960-1452   \\
200  &  1113-1284   &  1081-1338   &  950-1458   \\
400  &   998-1313   &   948-1351   &  938-1465   \\
800  &   665-1373   &   713-1415   &  722-1504   \\
\enddata

\tablenotetext{a}{{Temperatures in the $\bar{U}=0$ models isolate the
contribution from imposed day-night radiative forcing.  The bottom
surface of the model layer is bent by an amount corresponding to a
day-night temperature range of $(1\pm \eta)\times 1200$K
\citep[see][for details]{Cho2006}.  Trivial $\bar{U}=0$ models are
shown for reference only.}}
\end{deluxetable}

\clearpage

\end{document}